\documentclass[article]{llncs}
\usepackage{graphicx}
\usepackage{multirow}
\begin{document}
\title{iOCR: Informed Optical Character Recognition for Election Ballot Tallies}
%

%
\author{Kenneth U. Oyibo \and Jean D. Louis \and Juan E. Gilbert}

\authorrunning{K. Oyibo et al.}
%
\institute{University of Florida, Gainesville FL 32611, USA}
%
\maketitle              
\begin{abstract}
Integrity and trust in the election system are vital to the preservation of democracy. However, Ballot Marking Devices (BMD) used with barcode encoded votes raise concerns of transparency and security. The concern is that the official vote encoded in the barcode may not match the voter's intent. Furthermore, barcodes are not human-readable; therefore, vote manipulation would be undetectable by voters. One alternative to barcodes or Quick Response (QR) codes is Optical Character Recognition (OCR). Unfortunately, the character and spelling errors introduced by normal OCR systems are not suitable for an election. The purpose of this study is to explore the performance of Informed OCR (iOCR), which was developed with a spell correction algorithm to fix errors introduced by conventional OCR for vote tabulation. In this study, pre-tabulated ballots were scanned and then tabulated with a conventional OCR algorithm and iOCR. The results were compared based on vote tally accuracy and the number of spelling errors encountered. We also look at how the image quality of the ballot affects the tally produced by both systems. iOCR system outperforms conventional OCR techniques overall. The use of iOCR allows voters to simultaneously keep the accessibility and speed of a BMD while removing the barcode related concerns.
\end{abstract}

\section{Introduction}
Democracy hinges on the will of the voters. The process and technologies involved in voting have evolved tremendously, from in-person by-hand voting to modern Ballot Marking Devices (BMD). BMDs allow users to select their voting options on a screen and have the ballot printed out from a computer with the human-readable selectionS and a barcode of the encoded votes for the tally machine to use \cite{Appel2020}.

Introducing new technology has made elections more accessible and streamlined, but it has also brought new concerns. Using barcodes on ballots from BMDs has raised questions of risk from the cybersecurity community \cite{kieseberg2012}. One risk is a ``bad actor" manipulating the barcode's information so that it does not match the voter's intent. The voter can review the ballot's written portion but not the actual encoded data used for the tally. Nagy et al. state that coded representations must be interpretable by voters so that malicious software would not be able to print and encode different data to alter an election \cite{DIA}. A potential solution is to move away from barcodes and use Optical Character Recognition (OCR). Using OCR to tally the votes ensures that the user sees the exact information the system uses in the tally. OCR has not received widespread adoption due to inaccuracies introduced when recognizing characters. Consider two candidates, Cor\textbf{\itshape d}elia and Cor\textbf{\itshape n}elia running for the same position, and the OCR algorithm occasionally misrecognize the `d' or `n' in the candidates' names. The addition or removal of characters can change the outcome of an election.

There is a need to accurately count human-readable ballots and safeguard against errors. One solution is to use existing Spelling correction algorithms. Spelling correction algorithms are not new, but many are not well suited for the election setting. Some of these algorithms use word frequency dictionaries or precompiled lexicon of English words, not candidates' names \cite{norvig_2007}. What remains to be explored in the field is how to effectively use OCR for ballot tabulation from BMD. To address this gap, we present Informed OCR (iOCR). The iOCR system uses a combination of edit distance-based matching algorithms, per contest lexicons to address the ballots' spelling issues, candidate id, and confusion fail-safe to mitigate potential recognition errors. The addition of candidate id to the standard ballot format given by the National Institute of Standards and Technology (NIST) is in Fig.~\ref{figballot}. These methods are described in more detail in later sections.

The purpose of this study was to compare the accuracy of OCR to iOCR. We investigated the following research question: does the developed iOCR augmentation outperform regular OCR for vote tabulation. In this study, standard OCR and iOCR are given prefilled ballots to tally. The performance of each algorithm is then compared for vote count accuracy and the number of spelling errors encountered. We also investigated how image quality affects tabulation results. We hypothesized that the methods used in iOCR would correct common OCR errors and improve the accuracy of the vote tally results. The developed iOCR technique will eliminate the barcode concerns in our elections. 

\section{Background}
\subsection{Ballot Marking Devices}
A Ballot Marking Device (BMD) is an electronic voting device with a user interface that allows voters to select their choices. BMD provides different accessibility features, including but not limited to touchscreen interfaces, buttons, audio output, and font size adjustment \cite{Appel2020}. In 2020 United States general elections, BMDs were used in 47 out of the 50 states for in-person voting \cite{ballotpedia}. A BMD prints a paper version of the electronically marked ballot. Additionally, some BMDs print a barcode or QR code on the paper ballot along with a summary of the user's votes. The barcode or QR code represents a digital encoding of the voter's selections, which the optical scanner of a tally machine can interpret.

\begin{center}
\begin{figure}
\includegraphics[scale = 0.3]{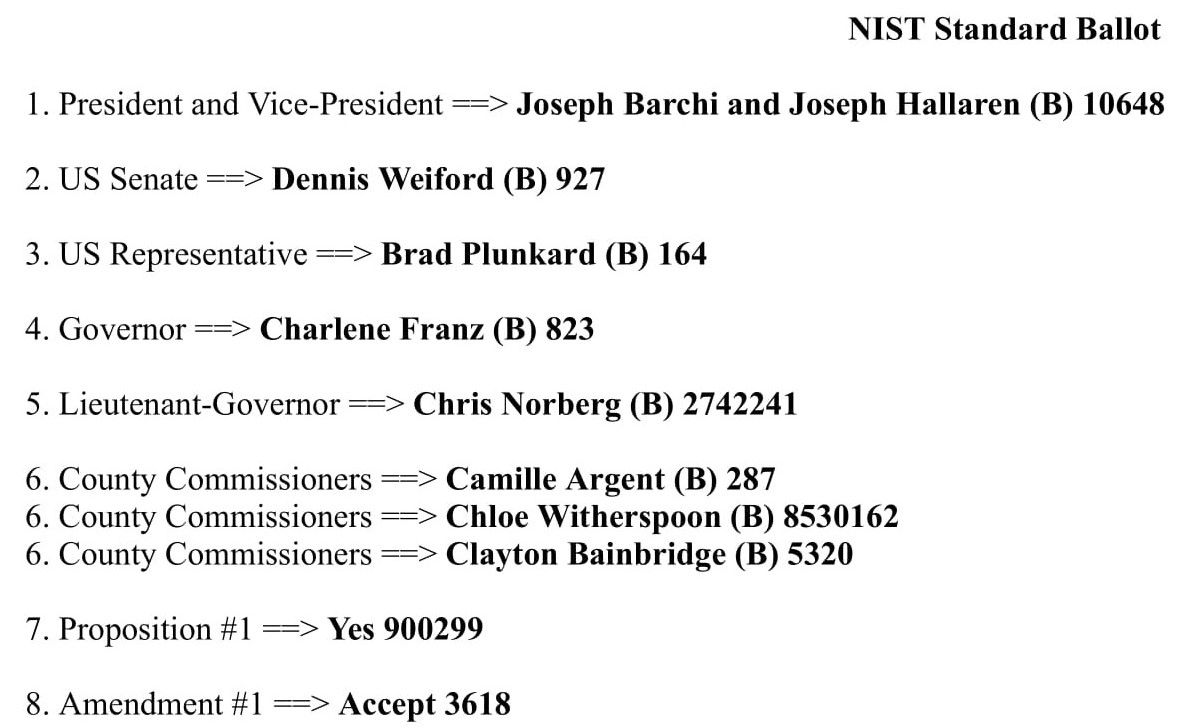}
\caption{A sample ballot, with added id}
\label{figballot}
\end{figure}
\end{center}

\subsection{Potential Vulnerabilities with Bar and QR codes}
Using barcodes and QR codes introduces potential vulnerabilities in elections. A hacked BMD may produce a QR code with malicious software encoded to manipulate the tally machine.
In an inconsistent barcode attack, the barcode is manipulated to be different from the printed human-readable summary \cite{wallach2020}. A voter may vote for one candidate, but a different candidate is encoded in the barcode. Therefore, while the user can verify the printed text, they cannot deduce whether the barcode or QR code is valid since the encoding is not human-readable \cite{kieseberg2012}.
\subsection{Optical Character Recognition}
Optical Character Recognition is a method of converting scanned images into machine-readable and editable text. The tool that does the actual conversion is known as the OCR engine, and several exist with varying performance metrics and error rates \cite{mithe_indalkar_divekar_2013}. The error rate of OCR engines is used to measure their performance. One of the significant limitations of OCR applications is that it is almost impossible to achieve 100\% accuracy. In a 2013 study, the English language character error rate and the word error rate of the Tesseract OCR were 0.47\% and 6.4\%, respectively \cite{smith2013}. The error rate is usually proportional to the quality of the input. Therefore, OCR is usually supplemented by preprocessing, which aims to improve the quality of the image before conversion. Post-processing targets correcting misrecognition, usually using a dictionary and an edit distance algorithm. Preprocessing involves techniques such as the removal of noise from the image and skew corrections. Post-processing uses a string similarity algorithm like Levenshtein distance and a dictionary derived from a large corpus of text to detect and correct misspellings in the OCR output text \cite{karthick2019steps}. In this study, we apply preprocessing with contest dictionaries as well as post-processing edit distance algorithms. We also test how image quality affects the tally results.

\subsection{Related Works}
The concept of correctly recognizing names from a document scanned by OCR has been crucial to digital text retrieval of books and other fields \cite{bui2017selecting}. This section presents three high-level methods used in other research for OCR improvement: consensus sequence voting OCR, preprocessed OCR, and post-processed OCR for mitigating OCR errors.
Authors in \cite{lopresti199} used voting amongst multiple OCR algorithms to correct OCR errors. They state that between 20\% and 50\% of the errors caused by an OCR system could be improved by scanning the document three times and running a `consensus sequence' voting procedure. This method's limitations are that it cannot correct errors when all the OCR voters disagree or all the voters produce the same misrecognition.

Bui et al. tackled OCR performance by investigating preprocessing techniques \cite{bui2017selecting}. They created synthetic distortions to test OCR. They concluded that the effectiveness of preprocessing methods depends on the "type of distortion" of the text, and sometimes the preprocessing worsens the performance of the OCR. Then they proposed a machine learning technique to select which type of preprocessing to do from 6 options, such as image sharpening or noise reduction.

In \cite{hamdi2019analysis} Hamdi et al. discuss OCR post-correction methods to extract textual content of digitised documents. One form of OCR post-correction is edit distance-based spelling correction which selects a word in the dictionary with the lowest distance to the misspelled word as the correct spelling. Hamdi et al. in \cite{hamdi2019analysis} have done work to study how OCR errors affect Name Entity Recognition (NER). They looked at how OCR post-correction affects the performance of NER from the noisy text. They used the Tesseract OCR library to read the file. They then used Levenshtein distance as the edit distance algorithm. Levenshtein distance calculates the minimum single character edits such as insertion, deletion, or substitution to transform one string to another string against a dictionary to make corrections. This method relatively fixes the voter consensus issues from the first method by having a dictionary to compare. The dictionary provides a ground truth as to what the text could be. The downside to this OCR correction method is that the algorithm sometimes changes a correct word in context with an incorrect one also found in a dictionary. The example they gave was that the algorithm might change  `Al-ain', a location name entity, with the word `Again'.

\section{Spell-Check Algorithms}
In this study, we explored some spell-checking implementations used to correct misspellings. These implementations rely on edit distance algorithms such as Levenshtein distance and Jaro-Winkler. The intent was to use these implementations to develop a strategy for correcting OCR misspellings. The result of the study carried out on these algorithms revealed their various advantages and disadvantages. Ultimately, it was decided that it was better for us to develop our solution using Levenshtein distance and Jaro-Winkler: informed OCR (iOCR) because the other implementations did not fit into the context of how we intend to use OCR in elections. The spelling algorithms investigated include: Peter Norvig's, Symspell, and deep learning algorithms.

\subsection{Peter Norvig Algorithm}
This algorithm works by generating a set of all possible edits up to the k edit distance of a misspelled word and selecting the likely correct spelling from the set where k can be any integer greater than or equal to zero. An edit to a word can be a deletion (remove letters), a transposition (swap two adjacent letters), a replacement (change one character to another, or an insertion (add extra letter(s)). The algorithm comprises 4 models: Selection model, Candidate Model, Language Model, and the Error model. The selection model selects the word with the highest frequency (occurrence) from the dictionary. The candidate model is responsible for generating all the possible edits of the misspelled word to be corrected. The language model approximates the probability of a word by enumerating the number of times it occurs in a large text file (usually a million words). The error model ranks possible correct candidates by edit distances and selects a lower edit distance as more probable \cite{norvig_2007}.

\subsection{Symspell}
This algorithm generates terms with an edit distance (deletes only) from each dictionary term and adds them together with the dictionary's original term. This addition must be done only once during a pre-calculation step. This algorithm is six orders of magnitude faster than Peter Norvig's because it uses only delete operations instead of transposes, replaces, inserts, and delete \cite{garbe_2014}.

\subsection{Deep Learning}
This technique is possible by combining text vectorization and deep learning. Text vectorization involves processing input strings into numerical representations. This technique allows mathematical functions to be carried out on words (i.e., words with similar or close edit distances can be represented within the same vector space). The vector representation can then be understood by deep neural networks using cosine similarity, which measures the similarity between two vectors of an inner product space.

The pros and cons of the different algorithms outlined above are summarized in the table below:

\begin{center}
\begin{table}[ht]
\caption{Spell-Check Algorithm Comparison}
\label{table2}
    \begin{center}
    \begin{tabular}{|c|c|c|}
    \hline
    Implementation & Pros & Cons \\
    \hline
    Peter Norvig & 
    \begin{tabular}[c]{@{}c@{}}Simple to use\\and implement\end{tabular} &
    \begin{tabular}[c]{@{}c@{}}Computationally\\expensive\\\hline Needs  frequency dictionary \\ of words\end{tabular}\\
    \hline
    Symspell & 
    \begin{tabular}[c]{@{}c@{}}Six orders of magnitude\\faster than\\Peter Norvig's \end{tabular} &
    \begin{tabular}[c]{@{}c@{}}Difficulty with\\special characters\end{tabular}\\
    \hline
    
    \begin{tabular}[c]{@{}c@{}}Deep\\Learning \end{tabular} & 
    \begin{tabular}[c]{@{}c@{}}Understands\\semantic context\end{tabular} &
    \begin{tabular}[c]{@{}c@{}} Difficulty with\\strings with spaces\\\hline Retraining and validation\\for each ballot\end{tabular}\\
    \hline
    
    \end{tabular}
    \end{center}
\end{table}
\end{center}

\section{Edit-Based Distance Algorithms}
After reviewing already existing solutions to use for our OCR post-processing ballot tallying, we decided to develop iOCR because they did not align with our context. The goal was to use edit-based distance algorithms to calculate string similarity scores between OCR string outputs and dictionary entries containing all possible voting choices. 

\subsection{Levenshtein Distance}
Levenshtein distance measures the similarity between two strings ($s_1, s_2$). It is the minimum number of characters insertion, deletion, and replacements needed to transform one string into another. Levenshtein similarity scales the scores to a range of 0 to 1, where 0 represents an exact match between both strings, and 1 represents no similarity between them \cite{naumann2013similarity}.

\begin{equation}
sim_{levenshtein} = 1 - \frac{LevenshteinDist}{max(|s_1|,|s_2|)}
\end{equation}

\subsection{Jaro-Winkler Similarity}
The Jaro-Winkler is a string similarity metric. Jaro-Winkler modifies the standard Jaro distance metric by putting extra weight on string differences at the start of the strings to be compared \cite{naumann2013similarity}. The metric is scaled between 0 and 1, where 0 represents no similarity, and 1 represents an exact match (opposite to Levenshtein scaling). When considering the context, if the start of the ballot line strings(context) are different, it is unlikely that it is a match.

\section{iOCR algorithm}
Informed OCR (iOCR) is an algorithm to complement regular OCR processing, which does not involve any form of post-processing to improve results. iOCR is included in the post-processing stage, and it is intended to fix any misrecognition contained in the string result of the ballot passed through an OCR engine. Tesseract OCR, a free, open-source OCR software managed by Google, is used. Pytesseract, a python package, is used as a wrapper for the Tesseract engine and provides the flexibility to process documents via python. We will refer to the Pytesseract simply as Tesseract OCR.
Two dictionary files containing all possible ballot selections for each contest are used. One of the dictionary files contains all possible ballot combinations with a candidate ID, and the other does not have candidate IDs. The dictionaries are used to calculate edit distances using string similarity algorithms between a possibly misrecognized ballot line and all the lines in the dictionary to find the exact match. The string similarity algorithms used are Levenshtein distance and Jaro-Winkler.

The iOCR algorithm starts by loading the dictionary file into different arrays. Lines belonging to the same election contest are stored in the same array. For example, lines containing \emph{President} and \emph{Vice-Presidents} will be in the same array, and lines containing \emph{Governor} will be grouped in another. This is implemented to optimize the accuracy and speed of the program by limiting the number of lines that the string similarity algorithms will have to calculate to find a match. The scanned ballot is passed through the Tesseract OCR engine, and a string output is returned. The program then removes any empty lines and extra garbage strings added in by Tesseract OCR. 

Afterward, the Levenshtein distance and Jaro- Winkler value of each ballot line in the Tesseract OCR string output is computed against each entry in the dictionary. The dictionary entries with the minimum Levenshtein distance and the maximum Jaro-Winkler value are selected as the possible correct matches. If these two selections are different, then it is an indication that we have two dictionary entries that are similar to the ballot line, but the algorithms cannot decide which one is the exact match. The ballot line and the two possible matches are written to a log file for human review. The algorithm also checks whether there are two or more dictionary entries with the same minimum Levenshtein or maximum Jaro-Winkler values. If this condition is true, then the ballot line is written to a log file for human review. However, if the two selections are the same, we can be confident that we found the correct match.

\begin{figure}
\includegraphics[scale = 0.1]{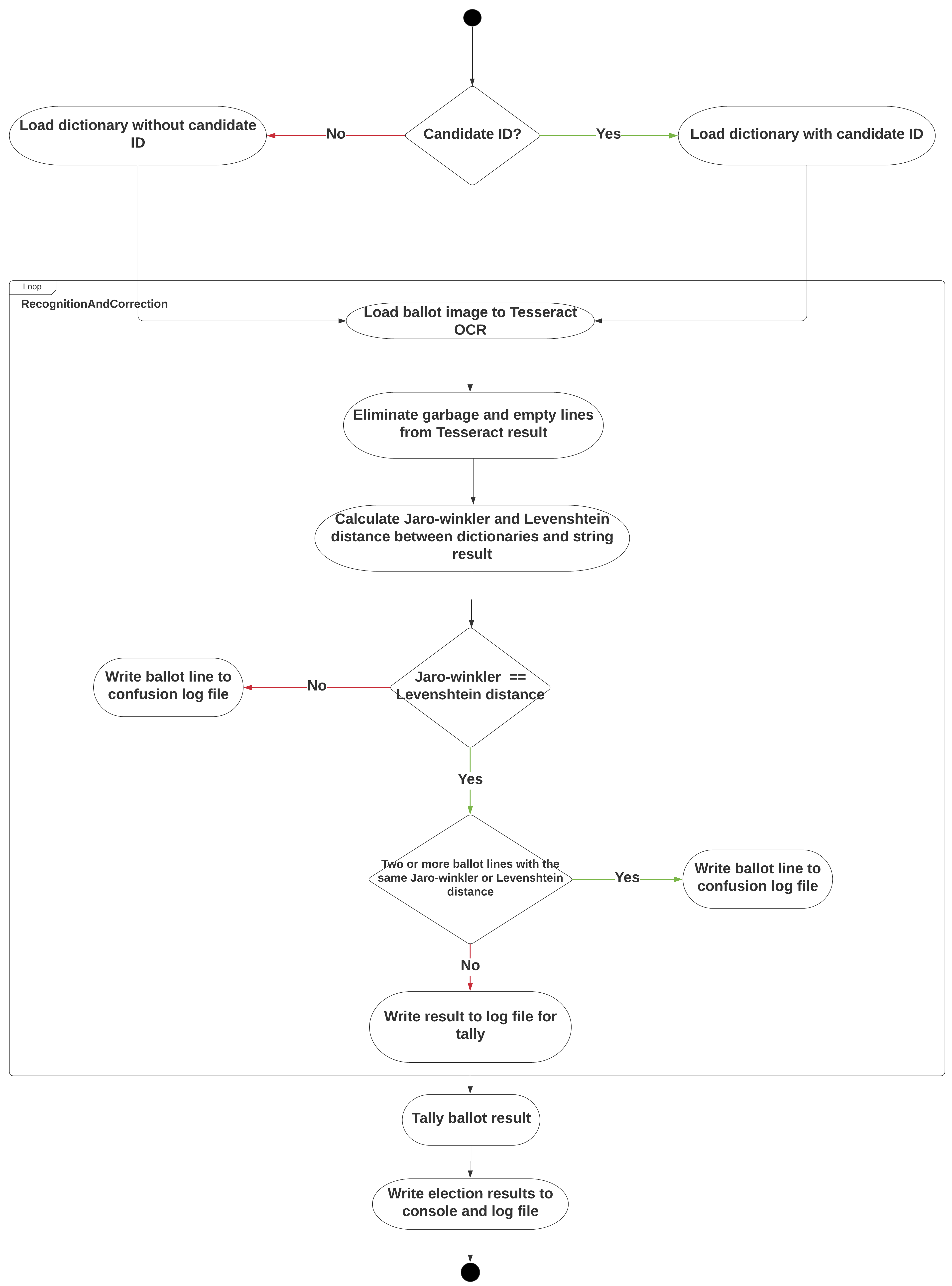}
\caption{iOCR Activity UML diagram}
\label{figuml}
\end{figure}

If we have found the correct match, the tally for the candidate for which the ballot was cast is updated. After all scanned ballot pages are processed, the total election result is tallied, and it is written to a log file and console. The activity flow of iOCR can be seen in ``Fig.~\ref{figuml}''.

\section{Experiments and Results}
Several experiments are conducted using election ballots from a mock election to test the performance and accuracy of the iOCR algorithm. The ballots are created using the Prime III software, an accessible electronic voting system developed in
\cite{smarr2017prime}. For these experiments, Prime III software, an open-source, accessible voting system, is used to generate two types of ballots: one that includes a unique candidate ID for each person and another without an id. The reason for doing this is to test whether including a unique candidate ID string of varying lengths will help edit distance algorithms differentiate similar strings better. ``Fig.~\ref{figballot}" is an example of a ballot with candidate id strings.
The performance of the iOCR is measured by checking if the tally at the end of the experiment matches the hand tallied result. There is an error if an entire line from the ballot is misrecognized, and the iOCR cannot find the correct match. The tally result of iOCR is compared to the tally of the regular Tesseract OCR engine without applying any form of post-processing.

\subsection{First Experiment Method}
The objective of this experiment was to compare the performance of the iOCR algorithm to a regular OCR. In this experiment, 50 ballots were created using the Prime III software. The names and party affiliations on the ballots were randomly selected from the sample of names and political parties already in the software. 25 of the ballots generated contained candidate IDs, and the other 25 did not contain a candidate ID. Both sets of the ballot are exactly the same except for the presence or absence of candidate IDs. In each of the 25 ballots, 5 of the ballots were write-in candidates containing names that were not anticipated by the dictionary. This is intended to replicate a real election where electorates might decide to vote for a person not on the ballot. Each ballot was duplicated 20 times, resulting in a total of 1000 ballots. The ballots are hand tallied to determine the correct result. The two ballot sets are processed through the iOCR program and standard OCR with no post-processing. The results are written to a log file for analysis and comparison with the hand tallied result.

Simple OCR accuracy was 0 because it misrecognized the first character of each line of the ballot (the contest number, shown in Fig \ref{figballot}). Since this was a consistent and fixable error, we disregarded this initial error. From comparing the results and the hand tallied results, the accuracy of the iOCR for both ballot set (with and without candidate ID) was 100\%. This suggests that the algorithms and techniques used in the iOCR were effective at correcting the string result from Tesseract. Compared to the hand-tallied result, the accuracy for the Tesseract OCR with no post-processing was also 100\% after disregarding the initial error.


\subsection{Second Experiment - Image Quality}
\begin{center}
\begin{figure}
\includegraphics[scale = 0.37]{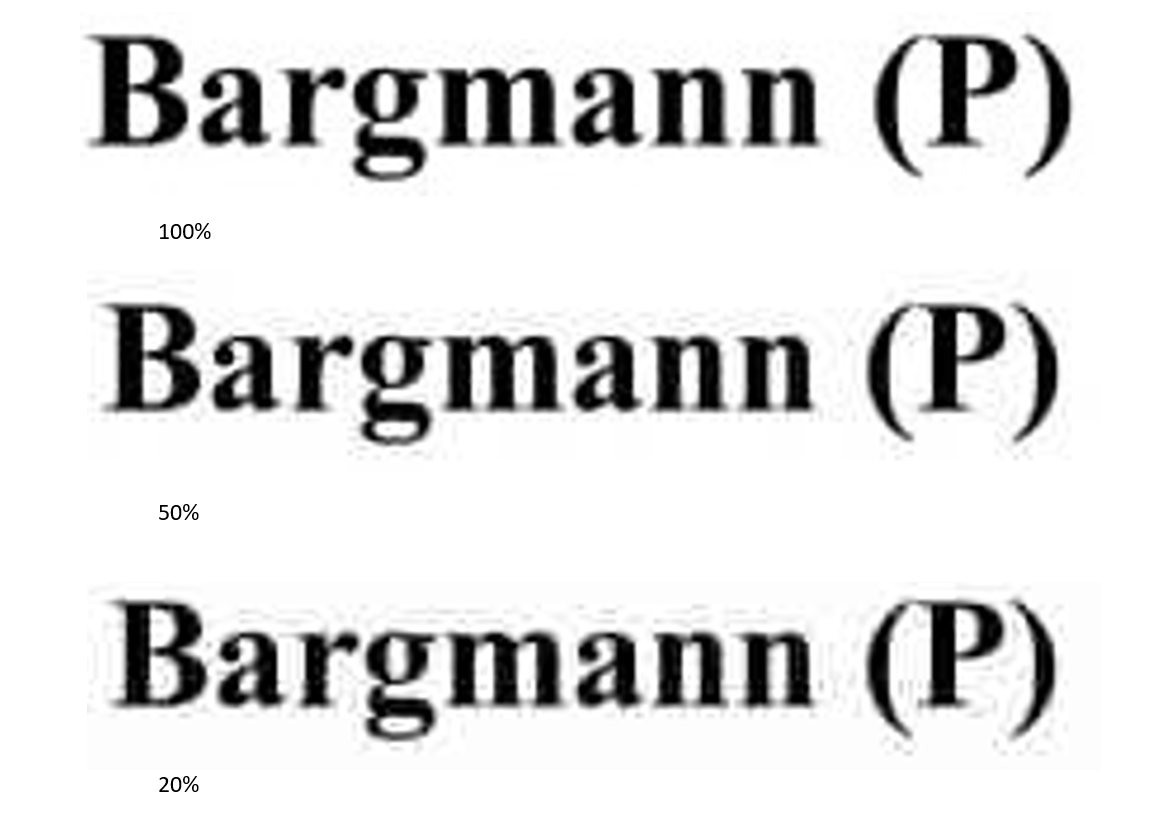}
\caption{500\% zoom at 100\%, 50\%, and 20\% quality}
\label{figzoom}
\end{figure}
\end{center}

Although Tesseract OCR gave the same degree of correctness as iOCR, there was skepticism that this was only true because of the quality of the image. We hypothesized that if the image quality were altered, Tesseract's character errors would be less consistent to the extent that tallying the result of an election will be difficult without post-processing. An election in Cuyahoga County, Ohio, found that manual counts differed from machines because 10 percent of ballots were degraded in some way, such as smeared, torn, or crumpled \cite{DIA}. Therefore, in this experiment, we reduced the quality of each ballot in the two-ballot set (with candidate ID and without) to 50\% and 20\% of the original quality. The original image is 2177x1560 pixels at 96dpi. The image quality reduction was made using python's PIL Image package \cite{clark}. The primary purpose of this experiment was to see the performance of the two systems as the quality of the images is reduced.
The ballot sets for both image quality were processed through the Tesseract OCR only and the iOCR. The program results were written to a log file for analysis and comparison with the hand tallied result. The degradation is not as visible until we zoom into the names as seen in ``Fig.~\ref{figzoom}''.

\subsubsection{Results}
For the image quality at 50\% dataset, iOCR was 100\% accurate for both the ballots sample with candidate ID and without. However, in the case of the Tesseract OCR with no post-processing, there were 60 lines in total out of 5000 ballot lines (500 ballots, each ballot has ten lines) that were inaccurate for both samples. Therefore, the accuracy was 98.8\%.
For the image quality at 20\% dataset, the iOCR was 100\% accurate for both ballot samples with candidate ID and without. In the Tesseract OCR test, 420 ballot lines out of 5000 lines were inaccurate in each of the two ballot samples. Therefore, the accuracy was reduced to 91.6\%.

A One-way ANOVA test using ballot accuracy as the dependent variable and image quality (100\%, 50\% and 20\%) group as the independent variable suggests that there is a significant difference among the three conditions (F(2,1497)= 143.43,p$<$0.0001) with ($\alpha$ = 0.05). Further T-test suggests that there is a significant difference between 100\% and 50\% (p $<$ 0.0001) and then 50\% and 20\% (p $< $0.0001).
Next, we compared the performance of OCR and iOCR using paired t-test with two image quality levels (50 and 20). There was a significant difference between the performance of OCR and iOCR accuracy at 20\% and 50\% image quality (t([499]=1.96), p $<$ 0.001 in both cases).

\subsection{Third Experiment Method  - iOCR Similarity Test}
This test evaluated how iOCR would perform for candidates with similar names that are only one edit distance apart that appear on the same ballot. Previous experiments had not considered this case. The candidate names ``Mark Day" and ``Mark May" were deliberately introduced into the Prime III software used to create the ballots. Our original data set did not contain this plausible scenario. Thirty ballots with Mark May as a candidate were created for the ballot sample with candidate IDs and without, then they were added to the existing ballot samples. Likewise, 20 ballots with Mark Day as a candidate were also added. The image quality was not altered or reduced in this experiment. Both names were included in the same race and party to make ballot lines containing both candidates very similar. This addition accounted for the worst-case scenario in terms of string similarity. 

\subsubsection{Results}
The experiment results showed that the accuracy was 100\% for both the ballot sample with candidate IDs and without one. However, further analysis of the log file containing the Tesseract OCR string result before post-processing revealed that the OCR never misrecognized the parts of the names. Therefore, it was inconclusive whether the iOCR will be accurate when either the "D" in Mark Day or the second "M" in Mark May are misrecognized for other characters or removed. As a result, we designed the fourth experiment where we intentionally misrecognized these characters.

\subsection{Fourth Experiment Method - iOCR Similarity Test}
This experiment evaluates how iOCR will perform when similar names are misrecognized. In this scenario, the two names are one edit distance apart, and the system misrecognized the differentiating letter. Instead of relying on Tesseract OCR to misrecognize the two names introduced, they are misspelled on purpose. Output ballot text containing various misspellings of ``Mark May" and ``Mark Day" are generated. The characters ``D" in ``Day" and ``M" in ``May" are replaced with other characters. One of the samples contains unique candidate IDs, while the other does not contain unique candidate IDs. Each ballot sample was passed through the iOCR program only, and the result is written to a log file for analysis and comparison with the hand tallied result.
\subsubsection{Results}
The accuracy of the iOCR test with ballots with candidate IDs was 100\%. However, for the iOCR test without candidate ID, each of the ballot lines that contained the misspelling was written to the confusion file because they were two potential candidates: the dictionary ballot lines of Mark Day and Mark May. This result was expected because the unique candidate IDs of varying lengths added to the ballot lines helped differentiate similar candidates. In this case, where there were two names that were only one edit distance different, the IDs set them apart.

\section{Discussion}
From the four experiments, we were able to make several observations. The Tesseract OCR is very consistent in its character recognition. This finding is because running the same image through the OCR engine is guaranteed to produce the same result no matter the number of times. This indicates that once a certain quality of image input that produces a satisfactory or accurate result text is achieved, the results will not vary arbitrarily unless the quality of input changes. Furthermore, Tesseract OCR was very accurate at character recognition. None of the characters were misrecognized when the image quality was at 100\%. The common character errors encountered with using Tesseract in our experiments were one or two space(s) insertion and deletion of characters. There were no character substitution errors.

When the image quality is at 100\%, the accuracy of iOCR and OCR are comparably the same. However, it should be noted that in our study for the Tesseract OCR, we adjusted the dictionary to match Tesseract output, missing the first characters. In a real-world application, this could lead to difficulty in tallying election results, especially if the location of the consistent character omission changes. Therefore, some form of post-processing, which iOCR does, is needed.

Although high-quality scanners that guarantee a high-quality image exist, it is quite possible that outliers, where the image quality is significantly lower, might come up. Therefore, iOCR is helpful in cases where the image quality might fluctuate. As observed in the results of experiment 2, iOCR was 100\% accurate both at 50\% and 20\% image quality, while Tesseract OCR was 98.8\% and 91.6\% accurate, respectively. The statistical test shows the change in accuracy to be significant and the difference between the Tesseract OCR and iOCR to be significant. iOCR outperforms Tesseract OCR.

The fourth experiment suggests that adding unique candidate IDs of varying length for each candidate effectively reduces the string similarity scores between candidates that are 1 edit distance apart. This ensures that no matter how similar two or more names are, iOCR with candidate IDs will most likely have a 100\% accuracy. For example, if Tom-1234 and Tim-43 are running for the same position and there is a misrecognition on the second letter, the algorithm would still be able to distinguish between Tim and Tom.

\subsection{OCR Quality}
Several factors affect the quality of an OCR output. One of them is the picture quality of the input image. In our second experiment, we observed that the quality of our OCR output decreased as the quality of the scanned image decreased. This finding is consistent with other research observations. It is recommended that the resolution be 300 dots per inch (dpi) and brightness of 50\% for best scanning results for OCR. Brightness settings that are too high or too low will decrease the accuracy of the image. In addition, the straightness of a scanned image affects the OCR quality. Skewed images will lead to more inaccurate recognition \cite{commons}.
Another limiting factor to consider is the candidate’s name similarity. Ballots that contain candidate names that are one edit distance apart will need to insert additional unique text to differentiate between them should the OCR misrecognize the character that sets them apart. Therefore, it is recommended that unique candidate IDs of varying lengths be added to ballot lines to reduce the string similarity between candidate names.
\subsection{Limitation}
The main limitations of this study are the sample size of ballots and candidate name distribution. We cannot predict which names will run for the same contest or how similar the names would be to test each possible scenario. The next step to overcome this limitation is to use iOCR to tabulate votes from an actual past election.

\section{Conclusion}
In this paper, we empirically explored the performance of iOCR compared to a standard OCR performance with four experiments. We found that iOCR is better suited for vote tally when compared to traditional OCR methods in certain cases. In the experiments, iOCR was either 100\% accurate or more accurate than the traditional method. iOCR accounts for errors in name similarity and character misrecognition and provides methods for resolving them. It also advances the field by implementing new methods to be robust to picture quality and errors. This work also highlights the effectiveness of candidate IDs in improving edit distance-based algorithms. 
Accuracy in ballot counting is crucial to an election. Fully human-readable paper ballots help confirm to the voter that what they see is what will be tallied. iOCR enables the continual use of accessible Ballot Marking Devices while eliminating one potential attack vector by removing the need for barcodes. Future work would include using a larger number of ballots with similar names and testing how other image quality degradation methods affect performance. We would like to further study the performance of iOCR to other standard OCR algorithms and test it on ballots from a current election.
\newpage
\section*{Acknowledgment}
This material is based upon work supported by the National Science Foundation Graduate Research Fellowship under Grant No. DGE-1842473. Any opinions, findings, and conclusions or recommendations expressed in this material are those of the author(s) and do not necessarily reflect the views of the National Science Foundation.

%
%
%
\bibliographystyle{splncs04}
\bibliography{refs}

\begin{thebibliography}{10}
\providecommand{\url}[1]{\texttt{#1}}
\providecommand{\urlprefix}{URL }
\providecommand{\doi}[1]{https://doi.org/#1}

\bibitem{Appel2020}
Appel, A.W., DeMillo, R.A., Stark, P.B.: Ballot-marking devices cannot ensure
  the will of the voters. Election Law Journal: Rules, Politics, and Policy
  \textbf{19}(3),  432--450 (9 2020)

\bibitem{ballotpedia}
Ballotpedia: Voting methods and equipment by state,
  \url{https://ballotpedia.org/Voting\_methods\_and\_equipment\_by\_state}

\bibitem{bui2017selecting}
Bui, Q.A., Mollard, D., Tabbone, S.: Selecting automatically pre-processing
  methods to improve ocr performances. In: 2017 14th IAPR International
  Conference on Document Analysis and Recognition (ICDAR). vol.~1, pp.
  169--174. IEEE (2017)

\bibitem{clark}
Clark, A.: python-pillow/pillow, \url{https://github.com/python-pillow/Pillow}

\bibitem{commons}
Commons, S.: Libguides: Introduction to ocr and searchable pdfs: Ocr best
  practices, \url{https://guides.library.illinois.edu/OCR/bestpractices}

\bibitem{garbe_2014}
Garbe, W.: Symspell (Mar 2014), \url{https://github.com/wolfgarbe/SymSpell}

\bibitem{hamdi2019analysis}
Hamdi, A., Jean-Caurant, A., Sidere, N., Coustaty, M., Doucet, A.: An analysis
  of the performance of named entity recognition over ocred documents. In: 2019
  ACM/IEEE Joint Conference on Digital Libraries (JCDL). pp. 333--334. IEEE
  (2019)

\bibitem{karthick2019steps}
Karthick, K., Ravindrakumar, K., Francis, R., Ilankannan, S.: Steps involved in
  text recognition and recent research in ocr; a study. International Journal
  of Recent Technology and Engineering  \textbf{8}(1),  2277--3878 (2019)

\bibitem{kieseberg2012}
Kieseberg, P., Schrittwieser, S., Leithner, M., Mulazzani, M., Weippl, E.,
  Munroe, L., Sinha, M.: Malicious pixels using qr codes as attack vector. In:
  Trustworthy ubiquitous computing, pp. 21--38. Springer (2012)

\bibitem{lopresti199}
Lopresti, D., Zhou, J.: Using consensus sequence voting to correct ocr errors.
  Computer Vision and Image Understanding  \textbf{67}(1),  39--47 (1997)

\bibitem{mithe_indalkar_divekar_2013}
Mithe, R., Indalkar, S., Divekar, N.: Optical character recognition (Mar 2013),
  \url{http://citeseerx.ist.psu.edu/viewdoc/download?doi=10.1.1.673.8061\&rep=rep1\&type=pdf}

\bibitem{DIA}
Nagy, G., Lopresti, D.: The role of document image analysis in trustworthy
  elections. In: Advances in Digital Document Processing and Retrieval, pp.
  51--81. World Scientific (2014)

\bibitem{naumann2013similarity}
Naumann, F.: Similarity measures. Information Systems  (2013)

\bibitem{norvig_2007}
Norvig, P.:  (Feb 2007), \url{https://norvig.com/spell-correct.html}

\bibitem{smarr2017prime}
Smarr, S.A., Sherman, I.N., Posadas, B., Gilbert, J.E.: Prime iii: Voting for a
  more accessible future. In: Proceedings of the 19th International ACM
  SIGACCESS Conference on Computers and Accessibility. pp. 335--336 (2017)

\bibitem{smith2013}
Smith, R.W.: History of the tesseract ocr engine: what worked and what didn't.
  In: Document Recognition and Retrieval XX. vol.~8658, p. 865802. SPIE (2013)

\bibitem{wallach2020}
Wallach, D.S.: On the security of ballot marking devices. Ohio St. Tech. LJ
  \textbf{16}, ~558 (2020)

\end{thebibliography}
%

\end{document}